\documentclass{Interspeech2024}
\usepackage{booktabs}
\usepackage[T1]{fontenc}
\usepackage{placeins}
\usepackage{bbding}
\usepackage{makecell}
\usepackage[utf8]{inputenc}

\interspeechcameraready

\title{MAT-SED: A Masked Audio Transformer with Masked-Reconstruction Based Pre-training for Sound Event Detection}
\name[affiliation={1}]{Pengfei}{Cai}
\name[affiliation={1}]{Yan}{Song}
\name[affiliation={1}]{Kang}{Li}
\name[affiliation={3}]{Haoyu}{Song}
\name[affiliation={2}]{Ian}{McLoughlin}

\address{
  $^1$National Engineering Research Center of Speech and Language Information Processing, University of Science and Technology of China, China
  $^2$ICT Cluster, Singapore Institute of Technology, Singapore 
  $^3$The Australian National University, Australia}
\email{cqi525@mail.ustc.edu.cn,
songy@ustc.edu.cn}


\begin{document}

\maketitle

\begin{abstract}
    

    Sound event detection~(SED) methods that leverage a large pre-trained Transformer encoder network have shown promising performance in recent DCASE challenges. 
    However, they still rely on an RNN-based context network to model temporal dependencies, largely due to the scarcity of labeled data. 
    In this work, we propose a pure Transformer-based SED model with masked-reconstruction based pre-training, termed MAT-SED.  
    Specifically, a Transformer with relative positional  encoding is first designed as the context network, pre-trained by the 
    masked-reconstruction task on all available target data in a self-supervised way. 
    Both the encoder and the context network are jointly fine-tuned in a semi-supervised manner. Furthermore, a global-local feature fusion strategy is proposed to enhance the localization capability. 
    Evaluation of MAT-SED on DCASE2023 task4 surpasses state-of-the-art performance, achieving 0.587/0.896 PSDS1/PSDS2 respectively.

\end{abstract}

\section{Introduction}
    Sound event detection~(SED) aims to recognize not only what events are happening in an audio signal but also when those events are happening.  
    Recent research in this field has garnered increasing interest from both academic and industrial sectors.
    The DCASE challenges~\footnote{\href{https://dcase.community/challenge2023/}{https://dcase.community/challenge2023/}} have been conducted to evaluate the performance of systems in environmental sound classification and detection, significantly driving the advancement of SED research.
    This technology is widely used in various applications, such as smart homes~\cite{krstulovic2018audio}, smart city~\cite{domazetovska2023iot}, surveillance~\cite{radhakrishnan2005audio}, etc.

\begin{figure}[t]
    \centering
    \includegraphics[width=\linewidth]{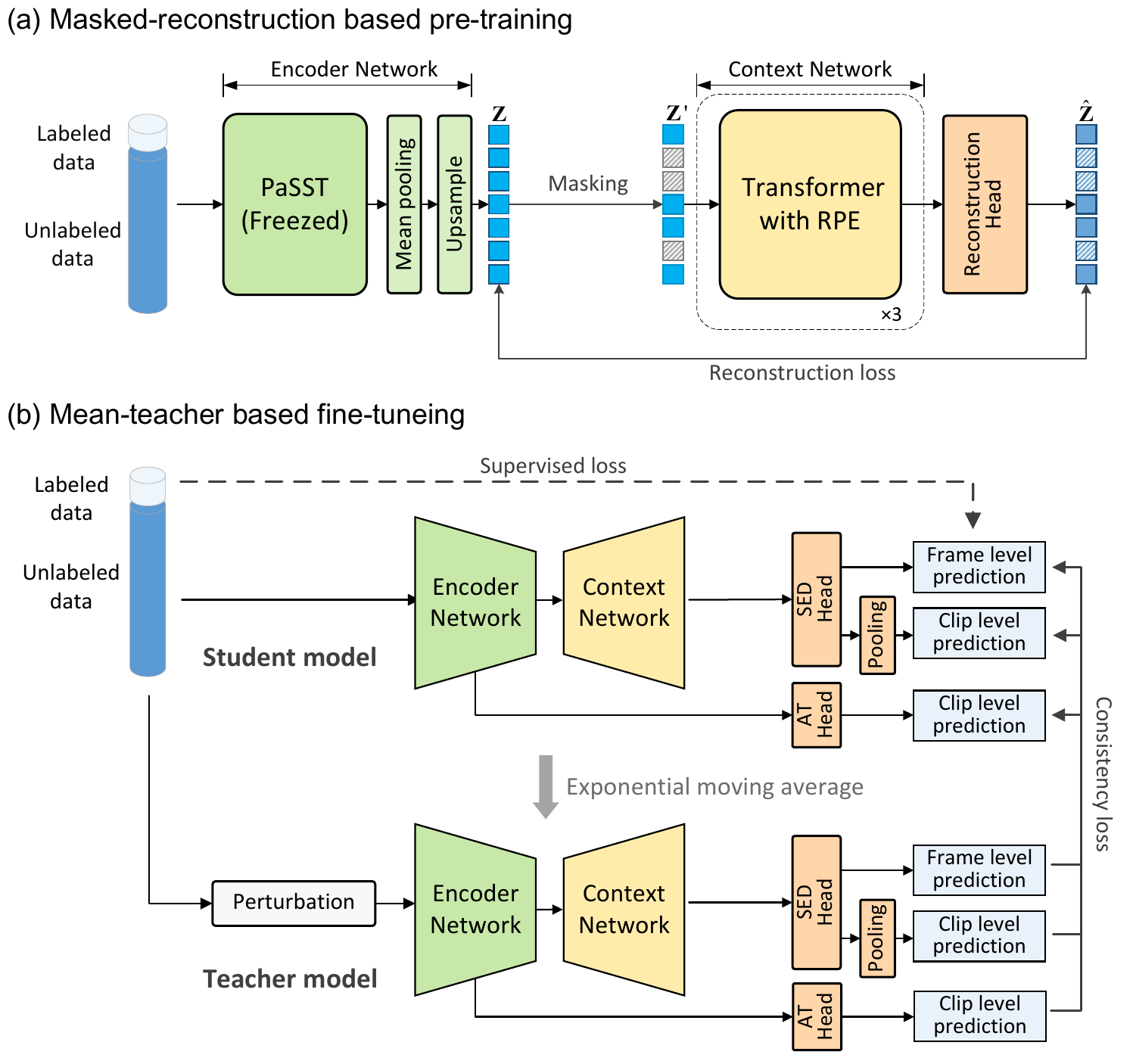}
    \caption{The architecture of MAT-SED, comprising two main components: the encoder network~(green) and the context network~(yellow), both of which are based on Transformer structures. "RPE" in the context network indicates the relative positional encoding.}
    \label{fig:framework}
    \end{figure}

     Most recent SED architecture can generally be divided into an encoder network and a context network, as illustrated in Figure \ref{fig:framework}.
     In classical CRNN based SED systems~\cite{cakir2017convolutional},  convolutional neural networks~(CNNs) are used as the encoder network for feature extraction, while recurrent neural networks~(RNNs) are employed as the context network to model temporal dependencies across latent features from the encoder.  
     The scarcity of labeled data is always a significant challenge for the SED task, due to the high cost of strong annotation for sound events.
     Semi-supervised methods, such as mean-teacher~\cite{tarvainen2017mean}, have thus been introduced to utilize large amounts of unlabeled data to mitigate the impact of insufficient labeled data.
   

     Recently, Transformer-based SED models have surged in popularity, inspired by the successes of Transformers in various domains, including  natural language processing~\cite{vaswani2017attention, DevlinCLT19}, computer vision~\cite{DosovitskiyB0WZ21} and automatic speech recognition~\cite{ZhangLSTMKK20, gulati20_interspeech}.  
     Convolution-augmented Transformer~\cite{Miyazaki2020} utilizes Conformer~\cite{gulati20_interspeech} instead of RNN to model temporal dependencies, winning the first place in DCASE2020 Task 4.
     That work demonstrated the potential of Transformer-based structures for SED, though performance was still limited due to insufficient labeled data.
     To mitigate the problem of data scarcity, a widely used approach is to employ Transformer models pre-trained  on readily available large-scale audio tagging datasets to serve as powerful feature extractors.
     Among high-ranking models~\cite{Kim2023, Zhang2023} of DCASE2023, the pre-trained Transformer and the CNN are concatenated in parallel as the encoder network, which can take the advantages of global and local features from different encoders.
     However, it is worth noting that most of those works only applied Transformer structures partially to the traditional CRNN.
     Again this is due to data scarcity issues. 
     Furthermore, although  powerful encoder networks can be obtained by pre-training, it is still difficult to train the downstream context network with limited labeled data.
     This remains a challenge to apply the pure Transformer-based structure for the SED task.  
     
    In this work, we present a pure Transformer-based SED model, termed Masked Audio Transformer for Sound Event Detection~(MAT-SED).
    MAT-SED begins with the pre-trained Transformer model as an encoder network, then a Transformer with relative positional encoding instead of RNNs as the context network, which can better capture long-range context dependencies of latent features.
    The Transformer structures lack some of the inductive biases inherent to RNNs, such as sequentiality, which makes the Transformer-based context networks do not generalize well when trained on insufficient data.
    To address this problem, we use the masked-reconstruction task to pre-train the context network in the self-supervised  manner, then fine-tune the pre-trained model with the classical mean-teacher algorithm. 
    This training paradigm maximizes the utilization of large quantities of unlabeled data compared to pure semi-supervised learning. 
    The global-local feature fusion strategy is employed to enhance the model’s localization accuracy in the fine-tuning stage. 
    Experimental results on the DCASE2023 dataset show that the proposed MAT-SED achieves 0.587/0.896 PSDS1/PSDS2, surpassing  state-of-the-art SED systems, thus demonstrating the potential of our approach.
    ~\footnote{The code is available at 
    \href{https://github.com/cai525/Transformer4SED}{https://github.com/cai525/Transformer4SED}}

\section{Methodology}
    In this section, we first outline the model structure of MAT-SED, then introduce the masked-reconstruction based pre-training and the fine-tuning strategies.

\subsection{Model}
    The overall structure of MAT-SED, as shown in Figure~\ref{fig:framework}, consists of two main components: the encoder network and the context network.
    The encoder network is used to extract features from the mel-spectrogram, outputting latent feature sequences.
    The context network is responsible for capturing temporal dependencies across the latent features. 
    Different types of head layer follow the context network to handle specific tasks, such as reconstruction, audio tagging and SED.

\subsubsection{Encoder network}
    The encoder network of MAT-SED is based on PaSST~\cite{koutini2021efficient}, a large pre-trained Transformer model for audio tagging. 
    Each mel-spectrogram is divided into several $16 \times 16$ patches, then patches are projected linearly to a sequence of embeddings.
    The sequence  traverses through 10 layers of PaSST blocks consisted of Transformers.
    Following PaSST, the frequency dimension is compressed via average pooling, succeeded by 10 times linear upsampling to restore the temporal resolution lost during the patching process.
    The output of the encoder network is denoted as $\mathbf{Z}=[\mathbf{z_1},\mathbf{z_2},...,\mathbf{z_T}]  \in \mathbb{R}^{C \times T}$, where C is the dimension of the embedding vector, and T is the length of encoder's output in the time dimension.

\subsubsection{Context network}
    Instead of the conventional RNN structure, we utilize 3 layers of Transformer block to constitute the context network. 
    Given the crucial need for localization in the SED task, integrating positional information becomes vital. 
    While RNN structures naturally embed positional information along the time dimension through their sequential structure~\cite{shaw-etal-2018-self}, Transformer models require positional encoding for the same purpose.
    The vanilla Transformer uses the absolute positional encoding~(APE)~\cite{vaswani2017attention, DevlinCLT19}, where the positional encoding depends on  absolute position of tokens.
    But for a given sound event, we hope that the model is translation equivariant along time dimension, i.e.  when the time of a sound event in an audio signal is changed, the same features will be detected at the new time.
    We therefore use relative positional encoding~(RPE)~\cite{dai2019transformer} to achieve this purpose, where the learnable positional encoding is determined by the relative position between frames.
    Compared to learnable APE, the RPE is naturally translation-equivariant~\cite{chu2023conditional}, making it more suitable for modelling temporal dependencies.

\begin{figure}[b]
  \centering
  \includegraphics[width=\linewidth]{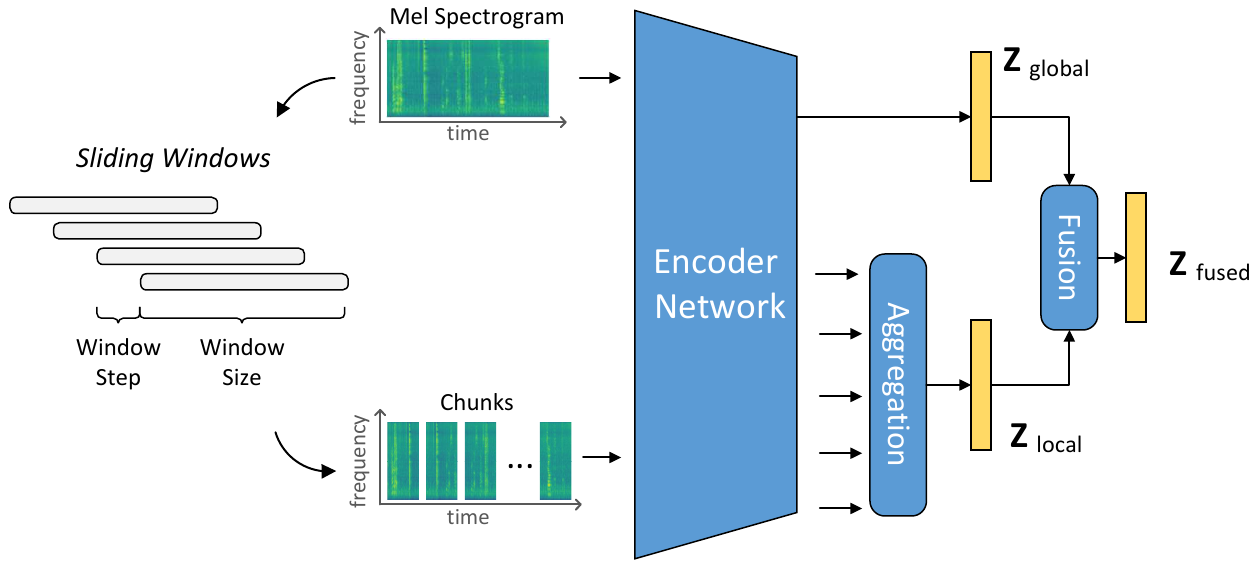}
  \caption{The global-local feature fusion strategy in the fine-tuning stage.
  }
  \label{fig:feature_map}
\end{figure}

\subsection{Masked-reconstruction based pre-training}
    The model structure during pre-training is depicted in Figure~\ref{fig:framework}~(a). 
    At this stage, we initialize the encoder network using the PaSST model pre-trained on AudioSet~\cite{gemmeke2017audio} and freeze its weights, to focus on pre-training the context network. 
    We design the masked-reconstruction task as the pretext task, similar to train a masked language model.
    We mask a certain proportion of frames in the latent feature sequence $\mathbf{Z}$, and substitute the masked frames with the learnable mask token, obtaining a new sequence $\mathbf{Z'}$.
    The masked-reconstruction task requires the context network to restore the masked latent features using the contextual information, which helps to enhance the temporal modeling ability of the context network.
    For the masking strategy, we adopt the block-wise masking strategy used in~\cite{bao2021beit}, dividing the sequence into several blocks of size 10, and masking entire blocks randomly.
    Compared to random masking, the block-wise masking strategy increases the difficulty of reconstruction, thus forces the model to learn more abstract semantic information. 
    The masked sequence traverses through the context network and the reconstruction head composed of two fully connected layers, yielding the reconstructed sequence $\mathbf{\hat{Z}}=[\hat{\mathbf{z}}_1,\hat{\mathbf{z}}_2,...,\hat{\mathbf{z}}_T] \in \mathbb{R}^{C \times T}$.We use mean squared error~(MSE) loss to evaluate the  quality of reconstruction:
    \begin{equation}
    \mathcal{L}_m = \sum_{x \in \mathcal{D}}\sum_{t \in M_x} (\hat{\mathbf{z}}_t(x) - \mathbf{z}_t(x))^2 \label{1} 
    \end{equation}
    where $\mathcal{D}$ denotes the pre-training dataset, and $M_x$ denotes the set of masked frame indices corresponding to the sample $x$.
    Note from this that only the masked frames are used to calculate the reconstruction loss.

\begin{table*}[t]
    \footnotesize
    \centering
    \caption{Comparison with state-of-the-art SED systems (~$^\dag$ denotes external data is used besides the DCASE2023 dataset).}
    \renewcommand\arraystretch{1.4}
    \begin{tabular}{m{3.5cm}<{\centering} m{3cm}<{\centering} m{2.5cm}<{\centering} m{1.5cm}<{\centering} m{1.5cm}<{\centering}}
    \Xhline{0.8pt}
        \textbf{Model} &   \textbf{Encoder Network}    &   \textbf{Context Network} & \textbf{PSDS1} & \textbf{PSDS2} \\
    \hline
        CRNN-BEATs       & Transformer + CNN  &  RNN   & 0.500 &  0.762           \\
        PaSST-SED~\cite{li23n_interspeech}  &Transformer      &  RNN   & 0.555 & 0.791    \\
        MFDConv-BEATs$^\dag$~\cite{Zhang2023}  & Transformer + CNN&  RNN   & 0.552  & 0.794             \\
        ATST-SED~\cite{shao2023finetune} & Transformer + CNN &  RNN   & 0.583 & 0.810             \\
    \hline
        MAT-SED~(median filter) &Transformer& Transformer   & \textbf{0.587} & 0.792         \\
        MAT-SED~(maximum filter) &Transformer& Transformer  & 0.090 &  \textbf{0.896}        \\
    \Xhline{0.8pt}
    \end{tabular}
    \label{tab1}
\end{table*}

\subsection{Fine-tuning}
    The model structure in the fine-tuning stage is shown in Figure~\ref{fig:framework}~(b).
    During fine-tuning, the reconstruction head is replaced by the SED head composed of a fully connected layer, which outputs the frame-level prediction.
    The frame-level prediction is pooled over the time dimension by linear-softmax pooling~\cite{wang2019comparison}, to obtain the clip-level prediction result.
    Following the task-aware module in~\cite{li23n_interspeech}, we additionally set up an AT head to focuse on the audio tagging task.
    The mean-teacher algorithm~\cite{tarvainen2017mean} is used for semi-supervised learning, with the consistency weight of 40. 

    Previous studies~\cite{Ebbers2022, gao23b_interspeech} have shown that using the window mechanism to limit the input duration can constrain the model to better attend to local information, thereby enhancing localization accuracy. 
    We thus propose a novel strategy for fine-tuning, termed global-local feature fusion strategy, as depicted in Figure~\ref{fig:feature_map}.  
    This strategy uses two branches to extract different features from the spectrogram. 
    The global branch feeds the original spectrogram into the encoder network,  yielding the global feature sequence $\mathbf{Z}_{global}$. In the local branch, the spectrogram is split into several overlapping chunks along the time dimension by sliding windows. Each chunk is then independently fed into the encoder network for feature extraction, and the output features from different chunks are aggregated to form the local feature sequence $\mathbf{Z}_{local}$.  Since there is temporal overlap between nearby chunks, we average the features extracted from different chunks for the overlapping duration. Global features $\mathbf{Z}_{global}$ and local features $\mathbf{Z}_{local}$ are fused linearly to obtain fused features $\mathbf{Z}_{fused}$:
    \begin{equation}
    \mathbf{Z}_{fused} = \lambda \mathbf{Z}_{local} + (1-\lambda)\mathbf{Z}_{global}
    \end{equation}
    In our model, $\lambda$ is set to 0.5 so fused features combine both local and global characteristics. It thus works well for sound events of varying duration.
    Since the overlap between chunks  increases the memory consumption significantly , the feature fusion strategy is only used in the teacher model of the mean-teacher algorithm, which does not require back-propagation.

\section{Experimental Setup}
\subsection{Dataset}
    The self-supervised pre-training and fine-tuning are both conducted on the DCASE2023~\cite{Turpault2019_DCASE} dataset, which is designed to detect sound event classes in domestic environments. 
    The training set consists of  10-second audio clips, including 1578 weakly-labeled clips, 3470 strongly-labeled clips, 10000 synthetic-strongly labeled clips, and 14412 unlabeled in-domain clips. The model is evaluated on the DCASE2023 challenge task 4 validation set, consisting of 1168 strongly-labeled clips. 

\subsection{Feature extraction and evaluation setting}
    The input audio is sampled at 32kHz. For feature extraction, we use a Hamming window of 25ms with a stride of 10ms to perform short-time Fourier transform(STFT). The spectrum obtained by the STFT is further transformed into a mel-spectrogram with 128 mel filters. Mixup~\cite{zhang2018mixup}, time shift and filterAugment~\cite{nam2022filteraugment} are used for data augmentation.

    The polyphonic sound detection score~(PSDS)~\cite{psds} is used as the evaluation metric.
    Following the setting of DCASE2023 competition, we use two different metric settings, PSDS1 and PSDS2, for two different scenarios. 
    The former focuses more on event localization, while the latter aims to avoid confusing between classes but for which localization is less crucial.
    Since PSDS1 can better reflect the model's localization performance, we use PSDS1 as the main evaluation metric in our experiments.
    In the testing phase, median filter and maximum filter are applied to the two PSDS scenarios respectively for post-processing~\cite{Li2023}.

\subsection{Model and training setting}
    For the sliding windows in the global-local feature fusion strategy, the window size and step are set to 5s and 0.3s. The context network contains 3 Transformer blocks with input dimension 768 , 12 attention heads, and expansion ratio 1.

    During the pre-training phase, the model is trained over 6000 steps with a batch size of 24 and a learning rate of $1\times 10^{-4}$.
    For the masked-reconstruction task, the masking rate is set to 75\%. During the fine-tuning stage, batch sizes for real strongly labeled, synthetic strongly labeled, real weakly labeled, and real unlabeled data are set to 3, 1, 4, 4, respectively.
    Following the strategy in ~\cite{baevski2020wav2vec}, only the SED head and AT head are trained for the first 6000 steps of fine-tuning, then the end-to-end fine-tuning is performed over the next 12000 steps.
    Learning rates for the encoder network, decoder network, and head layers are set to $5\times 10^{-6}$, $1\times 10^{-4}$, and $2\times 10^{-4}$, respectively.
    The AdamW~\cite{LoshchilovH19} optimizer is used for optimization with a weight decay of $1\times 10^{-4}$. Training is conducted on 2 Intel-3090 GPUs for 13 hours in total.

\section{Results}
     In this section, we first compare the performance of MAT-SED against other state-of-the-art SED models.
     Then, we conduct ablation experiments to analyze the contributions of each MAT-SED component.

\subsection{Performance of the proposed methods}
    Table~\ref{tab1} compares the performance of MAT-SED with other SED systems on the DCASE2023 dataset, where CRNN-BEATs is the baseline model of DCASE2023 task4.
    Our model achieves 0.587 PSDS1 and 0.896 PSDS2, outperforming previous SOTA models. It is noteworthy that MAT-SED stands out as the only model composed of pure Transformers in the table, whereas other models rely on CNN or RNN structures.
     This shows that the pure Transformer structure can perform well on SED tasks, given appropriate pre-training.

\subsection{Ablation studies}
\subsubsection{Ablations of the context network}

First, we explore the impact of different context network structures, as shown in Table~\ref{tab3}. Masked-reconstruction pre-training is employed in each set of experiments, and the hyperparameters of different structures are adjusted to the best. We use learnable APE in place of RPE to measure the effect of RPE. It can be seen from the table that the PSDS1 score of using RPE is significantly higher than APE, which indicates that the necessity of RPE for the SED task. Then we test the performance of Conformer ~\cite{gulati20_interspeech} for the context network. It can be seen from the table that  Conformer achieves a PSDS1 of 0.544,  trailing behind the Transformer using RPE, even though RPE is also utilized by Conformer. We suppose that the possible reason is that the convolution module in Conformer increases the parameter size, which makes it too bulky for the context network. Lastly, we substitute Transformers with GRU to compare the performance of RNNs and Transformers as the context network. The GRU achieves the PSDS1 of 0.557, lower than the Transformer using RPE, indicating that the Transformer with RPE serves as a more powerful context network structure than RNNs.

\begin{figure}[t]
  \centering
  \includegraphics[width=0.59\linewidth]{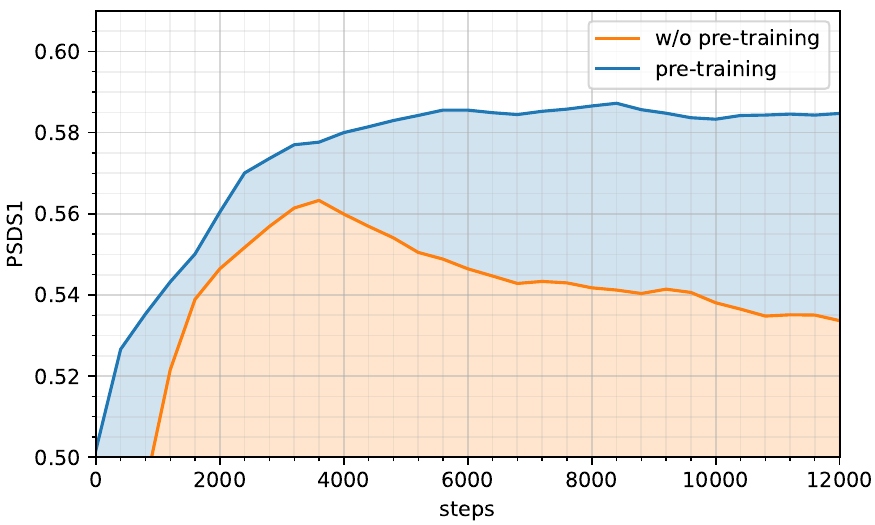}
  \caption{Convergence curves of training MAT-SED from scratch and end-to-end fine-tuning after masked-reconstruction pre-training.}
  \label{fig:pretrain_curve}
\end{figure}

\begin{table}[h]
    \footnotesize
    \centering
    \caption{Ablation study on the context network. "RPE" and "APE" denote relative positional encoding and absolute positional encoding respectively.}
    \renewcommand\arraystretch{1.2}
    \begin{tabular}{m{3cm}<{\centering} m{0.8cm}<{\centering} m{0.8cm}<{\centering} m{1cm}<{\centering}}
    \Xhline{0.8pt}
       \textbf{Context Network} & \textbf{RPE}&  \textbf{PSDS1} \\
    \hline
        Transformer with RPE& \Checkmark  & \textbf{0.587}\\
        Transformer with APE & \XSolidBrush  & 0.540\\
        Conformer& \Checkmark &0.544\\
        GRU& \XSolidBrush &0.557\\
    \Xhline{0.8pt}
    \end{tabular}
    \label{tab3}
\end{table}

\subsubsection{Ablations of masked-reconstruction based pre-training}
In this section, we analyze the effect of the masked-reconstruction pre-training. Figure~\ref{fig:pretrain_curve} compares the convergence curves of training MAT-SED from scratch and end-to-end fine-tuning after masked-reconstruction pre-training. For the pre-trained model, the SED layer and AT layer  are trained before the end-to-end fine-tuning to adjust to the features from pre-trained context network. The pre-trained network achieves a PSDS1 of 0.502 at the beginning of  end-to-end fine-tuning, even higher than the DCASE2023 baseline model, indicating that the representation learned by the context network in the masked-reconstruction pre-training is well-suited for the SED task. During the subsequent end-to-end fine-tuning process, the optimal PSDS1 score for the network without pre-training is 0.563, noticeably lower than the  pre-training network. On the other hand, severe overfitting occurs in the network without pre-training , which is not apparent in the pre-trained network. The results shows the efficacy of masked-reconstruction pre-training in enhancing Transformer-based context network's ability to model temporal dependencies, thus benefiting the localization of sound events.

\begin{figure}[tb]
  \centering
  \includegraphics[width=0.68\linewidth]{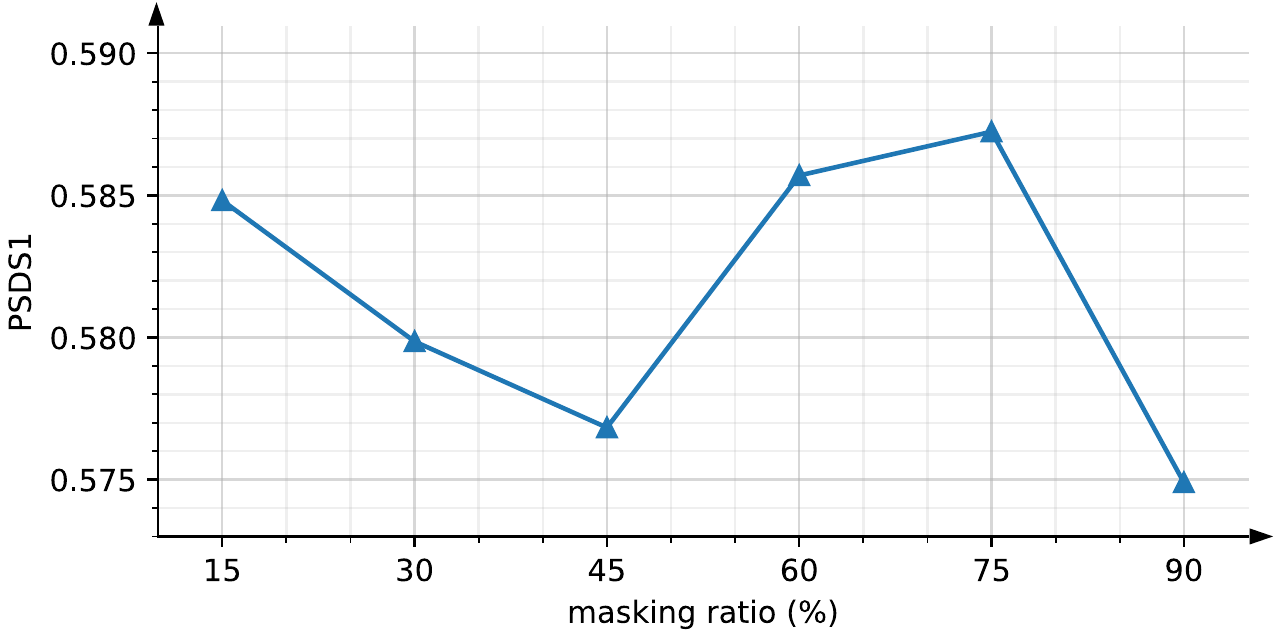}
  \caption{Impact of different masking ratio inimage the masked-reconstruction pre-training stage.}
  \label{fig:mask_rate_curve}
\end{figure}

Figure~\ref{fig:mask_rate_curve} further compares the effect of masking ratio in the masked-reconstruction pre-training. It can be seen from the figure that the optimal masking ratio is 75\%, relatively higher compared to Bert~(15\%)~\cite{DevlinCLT19}. A high masking ratio helps the model to learn abstract semantic features, rather than restoring the masked frames by simply interpolation. Similar conclusions have also been found in other self-supervised learning works based on masking \cite{baevski2020wav2vec, he2022masked}.

\begin{table}[t]
    \footnotesize
    \centering
    \caption{Ablation study on the hyperparameters $\lambda$ in \\the global-local feature fusion strategy.}
    \renewcommand\arraystretch{1.2}
    \begin{tabular}{m{1cm}<{\centering} |m{1cm}<{\centering} m{1cm}<{\centering} m{1cm}<{\centering}}
    \Xhline{1pt}
        $\lambda$ &   0 & 0.5 & 1 \\
    \hline
         PSDS1& 0.565 & 0.587 & 0.579\\
    \Xhline{0.8pt}
    \end{tabular}
    \label{tab2}
\end{table}

\subsubsection{Ablations of the  global-local feature fusion strategy}
In this section, we  analyze the effect of the  global-local feature fusion strategy in the fine-tuning stage. 
In the feature fusion strategy, the hyperparameter $\lambda$ controls the proportion of the global and local branches in the fused features. When $\lambda=1$, only  local features are retained; when $\lambda=0$, only  global features are retained, which means that the sliding window mechanism no longer works. In Table~\ref{tab2}, we compare the PSDS1 scores  when $\lambda$ is set to 0, 0.5 and 1. The experimental results show that higher PSDS1 is achieved when $\lambda=0.5$ than the cases when $\lambda$ is set to 0 or 1, indicating that fusing the global and local features can obtain more powerful latent features than only relying on either side.

\section{Conclusion}
In this paper, we propose MAT-SED, a pure Transformer-based SED model. 
In MAT-SED, the Transformer with relative positional encoding is employed as the context network, which enables the model to capture long-range context dependencies.
The masked-reconstruction task is used to pre-train the Transformer-based context network before semi-supervised based fine-tuning. The global-local feature fusion strategy is employed to further enhance the model’s localization accuracy.
MAT-SED achieves advanced performance on DCASE2023 dataset, outperforming other state-of-the-art SED models. Ablation experiments show that the self-supervised pre-training is crucial for Transformer-based structures.
In the future, we aim to further explore  self-supervised learning methods for audio Transformer structures.

\clearpage
\FloatBarrier
\bibliographystyle{IEEEtran}
\bibliography{wpref}

\begin{thebibliography}{10}
\providecommand{\url}[1]{#1}
\csname url@samestyle\endcsname
\providecommand{\newblock}{\relax}
\providecommand{\bibinfo}[2]{#2}
\providecommand{\BIBentrySTDinterwordspacing}{\spaceskip=0pt\relax}
\providecommand{\BIBentryALTinterwordstretchfactor}{4}
\providecommand{\BIBentryALTinterwordspacing}{\spaceskip=\fontdimen2\font plus
\BIBentryALTinterwordstretchfactor\fontdimen3\font minus
  \fontdimen4\font\relax}
\providecommand{\BIBforeignlanguage}[2]{{%
\expandafter\ifx\csname l@#1\endcsname\relax
\typeout{** WARNING: IEEEtran.bst: No hyphenation pattern has been}%
\typeout{** loaded for the language `#1'. Using the pattern for}%
\typeout{** the default language instead.}%
\else
\language=\csname l@#1\endcsname
\fi
#2}}
\providecommand{\BIBdecl}{\relax}
\BIBdecl

\bibitem{krstulovic2018audio}
S.~Krstulovi{\'c}, ``Audio event recognition in the smart home,''
  \emph{Computational Analysis of Sound Scenes and Events}, pp. 335--371, 2018.

\bibitem{domazetovska2023iot}
S.~Domazetovska, D.~Pecioski, V.~Gavriloski, and H.~Mickoski, ``Iot smart city
  framework using ai for urban sound classification,'' in \emph{INTER-NOISE and
  NOISE-CON Congress and Conference Proceedings}, vol. 265, no.~5.\hskip 1em
  plus 0.5em minus 0.4em\relax Institute of Noise Control Engineering, 2023,
  pp. 2767--2776.

\bibitem{radhakrishnan2005audio}
R.~Radhakrishnan, A.~Divakaran, and A.~Smaragdis, ``Audio analysis for
  surveillance applications,'' in \emph{IEEE Workshop on Applications of Signal
  Processing to Audio and Acoustics, 2005.}\hskip 1em plus 0.5em minus
  0.4em\relax IEEE, 2005, pp. 158--161.

\bibitem{cakir2017convolutional}
E.~Cak{\i}r, G.~Parascandolo, T.~Heittola, H.~Huttunen, and T.~Virtanen,
  ``Convolutional recurrent neural networks for polyphonic sound event
  detection,'' \emph{IEEE/ACM Transactions on Audio, Speech, and Language
  Processing}, vol.~25, no.~6, pp. 1291--1303, 2017.

\bibitem{tarvainen2017mean}
A.~Tarvainen and H.~Valpola, ``Mean teachers are better role models:
  Weight-averaged consistency targets improve semi-supervised deep learning
  results,'' \emph{Advances in neural information processing systems}, vol.~30,
  2017.

\bibitem{vaswani2017attention}
A.~Vaswani, N.~Shazeer, N.~Parmar, J.~Uszkoreit, L.~Jones, A.~N. Gomez,
  {\L}.~Kaiser, and I.~Polosukhin, ``Attention is all you need,''
  \emph{Advances in neural information processing systems}, vol.~30, 2017.

\bibitem{DevlinCLT19}
J.~Devlin, M.~Chang, K.~Lee, and K.~Toutanova, ``{BERT:} pre-training of deep
  bidirectional transformers for language understanding,'' in \emph{Proceedings
  of the 2019 Conference of the North American Chapter of the Association for
  Computational Linguistics}.\hskip 1em plus 0.5em minus 0.4em\relax
  Association for Computational Linguistics, 2019, pp. 4171--4186.

\bibitem{DosovitskiyB0WZ21}
A.~Dosovitskiy, L.~Beyer, A.~Kolesnikov, D.~Weissenborn, X.~Zhai,
  T.~Unterthiner, M.~Dehghani, M.~Minderer, G.~Heigold, S.~Gelly, J.~Uszkoreit,
  and N.~Houlsby, ``An image is worth 16x16 words: Transformers for image
  recognition at scale,'' in \emph{9th International Conference on Learning
  Representations (ICLR)}, 2021.

\bibitem{ZhangLSTMKK20}
Q.~Zhang, H.~Lu, H.~Sak, A.~Tripathi, E.~McDermott, S.~Koo, and S.~Kumar,
  ``Transformer transducer: {A} streamable speech recognition model with
  transformer encoders and {RNN-T} loss,'' in \emph{{IEEE} International
  Conference on Acoustics, Speech and Signal Processing (ICASSP)}.\hskip 1em
  plus 0.5em minus 0.4em\relax {IEEE}, 2020, pp. 7829--7833.

\bibitem{gulati20_interspeech}
A.~Gulati, J.~Qin, C.-C. Chiu, N.~Parmar, Y.~Zhang, J.~Yu, W.~Han, S.~Wang,
  Z.~Zhang, Y.~Wu, and R.~Pang, ``{Conformer: Convolution-augmented Transformer
  for Speech Recognition},'' in \emph{Proc. Interspeech 2020}, 2020, pp.
  5036--5040.

\bibitem{Miyazaki2020}
K.~Miyazaki, T.~Komatsu, T.~Hayashi, S.~Watanabe, T.~Toda, and K.~Takeda,
  ``Convolution-augmented transformer for semi-supervised sound event
  detection,'' DCASE2020 Challenge, Tech. Rep., June 2020.

\bibitem{Kim2023}
J.~W. Kim, S.~W. Son, Y.~Song, .~Kim, Hong~Kook1, I.~H. Song, and J.~E. Lim,
  ``Semi-supervised learning-based sound event detection using frequency
  dynamic convolution with large kernel attention for {DCASE} challenge 2023
  task 4,'' DCASE2023 Challenge, Tech. Rep., June 2023.

\bibitem{Zhang2023}
S.~Xiao, J.~Shen, A.~Hu, X.~Zhang, P.~Zhang, and Y.~Yan, ``Sound event
  detection with weak prediction for dcase 2023 challenge task4a,'' DCASE2023
  Challenge, Tech. Rep., June 2023.

\bibitem{koutini2021efficient}
K.~Koutini, J.~Schlüter, H.~Eghbal-zadeh, and G.~Widmer, ``{Efficient Training
  of Audio Transformers with Patchout},'' in \emph{Proc. Interspeech 2022},
  2022, pp. 2753--2757.

\bibitem{shaw-etal-2018-self}
P.~Shaw, J.~Uszkoreit, and A.~Vaswani, ``Self-attention with relative position
  representations,'' in \emph{Proceedings of the 2018 Conference of the North
  {A}merican Chapter of the Association for Computational Linguistics}.\hskip
  1em plus 0.5em minus 0.4em\relax Association for Computational Linguistics,
  Jun. 2018, pp. 464--468.

\bibitem{dai2019transformer}
Z.~Dai, Z.~Yang, Y.~Yang, J.~G. Carbonell, Q.~V. Le, and R.~Salakhutdinov,
  ``Transformer-xl: Attentive language models beyond a fixed-length context,''
  in \emph{Proceedings of the 57th Conference of the Association for
  Computational Linguistics, {ACL}}.\hskip 1em plus 0.5em minus 0.4em\relax
  Association for Computational Linguistics, 2019, pp. 2978--2988.

\bibitem{chu2023conditional}
X.~Chu, Z.~Tian, B.~Zhang, X.~Wang, and C.~Shen, ``Conditional positional
  encodings for vision transformers,'' in \emph{The Eleventh International
  Conference on Learning Representations}, 2023.

\bibitem{gemmeke2017audio}
J.~F. Gemmeke, D.~P. Ellis, D.~Freedman, A.~Jansen, W.~Lawrence, R.~C. Moore,
  M.~Plakal, and M.~Ritter, ``Audio set: An ontology and human-labeled dataset
  for audio events,'' in \emph{IEEE international conference on acoustics,
  speech and signal processing (ICASSP)}.\hskip 1em plus 0.5em minus
  0.4em\relax IEEE, 2017, pp. 776--780.

\bibitem{bao2021beit}
H.~Bao, L.~Dong, S.~Piao, and F.~Wei, ``{BEiT}: {BERT} pre-training of image
  transformers,'' in \emph{International Conference on Learning
  Representations}, 2022.

\bibitem{li23n_interspeech}
K.~Li, Y.~Song, I.~McLoughlin, L.~Liu, J.~Li, and L.-R. Dai, ``{Fine-tuning
  Audio Spectrogram Transformer with Task-aware Adapters for Sound Event
  Detection},'' in \emph{Proc. INTERSPEECH 2023}, 2023, pp. 291--295.

\bibitem{shao2023finetune}
N.~Shao, X.~Li, and X.~Li, ``Fine-tune the pretrained atst model for sound
  event detection,'' \emph{arXiv preprint arXiv:2309.08153}, 2023.

\bibitem{wang2019comparison}
Y.~Wang, J.~Li, and F.~Metze, ``A comparison of five multiple instance learning
  pooling functions for sound event detection with weak labeling,'' in
  \emph{IEEE International Conference on Acoustics, Speech and Signal
  Processing (ICASSP)}.\hskip 1em plus 0.5em minus 0.4em\relax IEEE, 2019, pp.
  31--35.

\bibitem{Ebbers2022}
J.~Ebbers and R.~Haeb-Umbach, ``Pre-training and self-training for sound event
  detection in domestic environments,'' DCASE2022 Challenge, Tech. Rep., June
  2022.

\bibitem{gao23b_interspeech}
L.~Gao, Q.~Mao, and M.~Dong, ``{Joint-Former: Jointly Regularized and Locally
  Down-sampled Conformer for Semi-supervised Sound Event Detection},'' in
  \emph{Proc. INTERSPEECH 2023}, 2023, pp. 2753--2757.

\bibitem{Turpault2019_DCASE}
N.~Turpault, R.~Serizel, A.~Parag~Shah, and J.~Salamon, ``{Sound event
  detection in domestic environments with weakly labeled data and soundscape
  synthesis},'' in \emph{{Workshop on Detection and Classification of Acoustic
  Scenes and Events}}, October 2019.

\bibitem{zhang2018mixup}
H.~Zhang, M.~Cisse, Y.~N. Dauphin, and D.~Lopez-Paz, ``mixup: Beyond empirical
  risk minimization,'' in \emph{International Conference on Learning
  Representations (ICLR)}, 2018.

\bibitem{nam2022filteraugment}
H.~Nam, S.-H. Kim, and Y.-H. Park, ``Filteraugment: An acoustic environmental
  data augmentation method,'' in \emph{IEEE International Conference on
  Acoustics, Speech and Signal Processing (ICASSP)}.\hskip 1em plus 0.5em minus
  0.4em\relax IEEE, 2022, pp. 4308--4312.

\bibitem{psds}
C.~Bilen, G.~Ferroni, F.~Tuveri, J.~Azcarreta, and S.~Krstulovic, ``A framework
  for the robust evaluation of sound event detection,'' in \emph{IEEE
  International Conference on Acoustics, Speech and Signal Processing
  (ICASSP)}, 2020, pp. 61--65.

\bibitem{Li2023}
K.~Li, P.~Cai, and Y.~Song, ``Li {USTC} team's submission for {DCASE} 2023
  challenge task4a,'' DCASE2023 Challenge, Tech. Rep., June 2023.

\bibitem{baevski2020wav2vec}
A.~Baevski, Y.~Zhou, A.~Mohamed, and M.~Auli, ``wav2vec 2.0: A framework for
  self-supervised learning of speech representations,'' \emph{Advances in
  neural information processing systems}, vol.~33, pp. 12\,449--12\,460, 2020.

\bibitem{LoshchilovH19}
I.~Loshchilov and F.~Hutter, ``Decoupled weight decay regularization,'' in
  \emph{7th International Conference on Learning Representations (ICLR)}, 2019.

\bibitem{he2022masked}
K.~He, X.~Chen, S.~Xie, Y.~Li, P.~Doll{\'a}r, and R.~Girshick, ``Masked
  autoencoders are scalable vision learners,'' in \emph{Proceedings of the
  IEEE/CVF conference on computer vision and pattern recognition}, 2022, pp.
  16\,000--16\,009.

\end{thebibliography}

\end{document}